# Wray 977 (GX301-2): a hypergiant with pulsar companion*

**L. Kaper[1], H.J.G.L.M. Lamers[2,3], E. Ruymaekers[4,5,6], E.P.J. van den Heuvel[4,5], and E.J. Zuiderwijk[7,8]**

[1] European Southern Observatory, Karl Schwarzschild Str. 2, D-85748 Garching bei München, Germany
[2] Sterrekundig Instituut, Princetonplein 5, Postbus 80000, Utrecht, The Netherlands
[3] SRON Utrecht, Sorbonnelaan 2, 3584 CA Utrecht, The Netherlands
[4] Astronomical Institute "Anton Pannekoek", University of Amsterdam, Kruislaan 403, 1098 SJ Amsterdam, The Netherlands
[5] Center for High Energy Astrophysics (CHEAF), Kruislaan 403, 1098 SJ Amsterdam, The Netherlands
[6] Instituut voor Sterrenkunde, K.U. Leuven, Celestijnenlaan 200B, 3001 Heverlee, Belgium
[7] Kapteyn Astronomical Institute, Postbus 800, 9700 AV Groningen, The Netherlands
[8] Royal Greenwich Observatory, Madingley Road, Cambridge CB3 0EZ, England

**Abstract.** Wray 977, the B supergiant companion of the X-ray pulsar GX301-2, should be classified as B1 Ia+, based on a comparison of its optical spectrum to that of $\zeta^1$ Sco, a well-known B hypergiant and one of the brightest stars in our galaxy. The classification of Wray 977 as a hypergiant results in a new distance determination of the binary system, i.e. $d = 5.3$ kpc (previously $1.8 \pm 0.4$ kpc). The "average" X-ray luminosity of the pulsar is then $\sim 10^{37}$ erg s$^{-1}$, in good agreement with the predicted X-ray luminosity resulting from accretion of a dense, low-velocity ($v_\infty = 400$ km s$^{-1}$) stellar wind. A mass-loss rate of $\lesssim 10^{-5}$ M$_\odot$ yr$^{-1}$ is estimated from the H$\alpha$ profile. A new upper limit for the inclination of the system is derived which provides a lower limit to the (present) mass of Wray 977 (48 M$_\odot$). Regarding current binary evolution scenarios, the empirical lower mass limit for black-hole formation in a binary increases to M$_{ZAMS} \gtrsim 50$ M$_\odot$.

## 1. Introduction

The most recent spectral study of Wray 977 (BP Cru), the optical counterpart (Vidal 1973) of the binary X-ray pulsar GX 301-2 (4U1223-62) was presented by Parkes et al.

(1980). They derived a spectral type of B2 Iae for the supergiant and concluded that the mass-loss rate is probably higher than $\sim 3 \times 10^{-6}$ M$_\odot$ yr$^{-1}$, adding the remark that "the spectrum of Wray 977 does show more signs of mass loss than a 'normal' B2 supergiant". Adopting $V = 10.83$ (Hammerschlag-Hensberge et al. 1976) and absolute magnitude $M_V = -7.0$ for a B2 Ia star (Balona & Crampton 1974), they derived a distance to Wray 977 of $1.8 \pm 0.4$ kpc. This distance implied an average X-ray luminosity of $\sim 10^{36}$ erg s$^{-1}$ for the 696s pulsar GX 301-2.

The orbital period of this high-mass X-ray binary system (HMXB) is 41.5 days (Watson et al. 1982), based on the periodicity in X-ray flaring events of GX 301-2. Pulse-timing analysis carried out by White & Swank (1984) and, from a more extended dataset, by Sato et al. (1986) resulted in the accurate determination of the orbital elements of GX 301-2. The system is characterized by a large eccentricity ($e = 0.47$), and the mass function $f(m) = 31.9 \pm 0.8$ M$_\odot$ is the highest known for a HMXB with pulsar companion. The latter fact was used by Van den Heuvel & Habets (1984) to derive an observational lower mass limit for black-hole formation ($M_{ZAMS} \geq 40$ M$_\odot$) in a massive binary. Given the supergiant's radius (43 R$_\odot$, Parkes et al. 1980), the absence of an eclipse in the X-ray lightcurve constrains the inclination of the system to $i \leq 78°$. A plausible set of parameters is $i \approx 75°$ and hence, from the mass function, M $\approx 38$ M$_\odot$ (Sato et al. 1986).

Several authors have tried to explain the observed modulation of the X-ray luminosity of GX 301-2 with orbital phase. White & Swank (1984) proposed that the mass loss from Wray 977 must be extreme and that the wind velocity at the neutron star's orbit is much lower than usually adopted for a normal B supergiant. Leahy

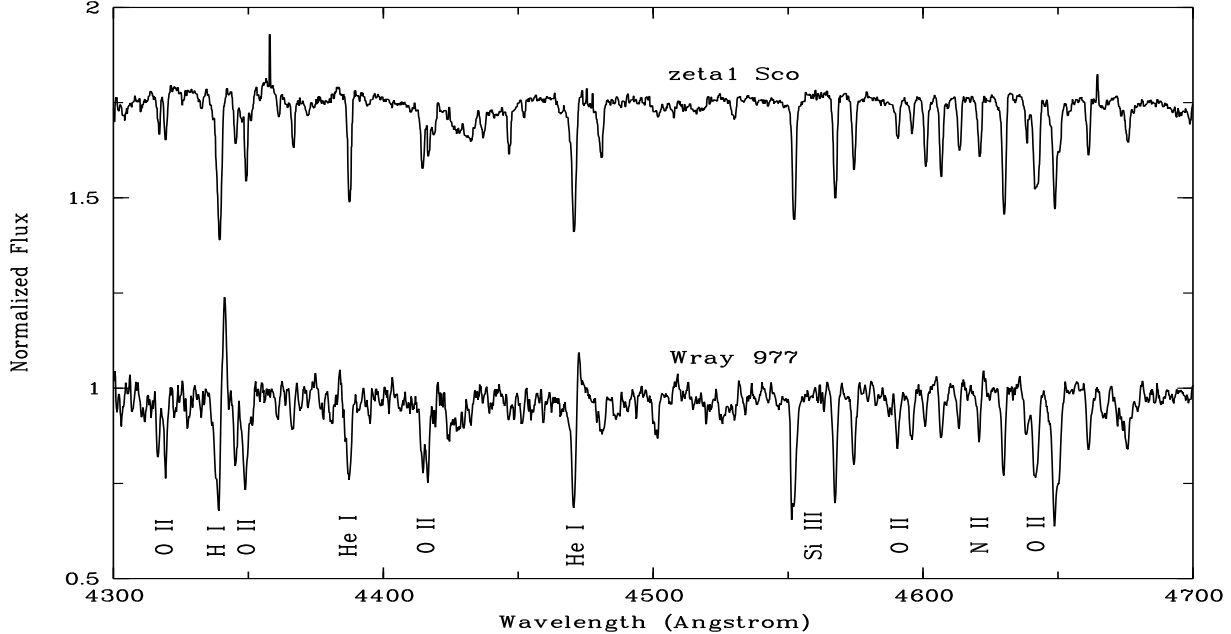

**Fig. 1.** The optical spectrum of Wray 977 (bottom) from 4300-4700 Å and that of the hypergiant $\zeta^1$ Sco (top) are very similar. The spectrum of $\zeta^1$ Sco was kindly provided to us by Dr. O. Stahl. The H$\gamma$ and He I 4471 Å line of Wray 977 are stronger and have a P Cygni-type profile suggesting that Wray 977 has an even higher mass-loss rate than $\zeta^1$ Sco. The O II lines are stronger and the N II lines weaker than those in the spectrum of $\zeta^1$ Sco, indicating that Wray 977 has a slightly earlier spectral type than $\zeta^1$ Sco, i.e. B1 Ia+. The strong O II lines, the relatively weak He I 4387 Å line in comparison with the Si III lines, and the strong Balmer emission are consistent with an extremely high luminosity of Wray 977

(1991) and Haberl (1991) suggested that a gas stream in the system might be responsible for the observed drastic increase in X-ray luminosity just before each periastron passage. In the following, we will present arguments that Wray 977 is not a normal B supergiant, but a B hypergiant. This results in a new distance determination of the system; new estimates for the various stellar parameters are given in Section 3, including a new lower limit to the mass of Wray 977. In the final section we discuss the implications for massive binary evolution.

## 2. Spectral classification and luminosity of Wray 977

In Figs. 1 and 2 we show parts of the optical spectrum of Wray 977, obtained in February 1984, with the Cassegrain spectrograph (CASPEC) mounted on the 3.6m telescope at the European Southern Observatory in Chile (Tab. 1). The spectra (partly) cover the wavelength region from 4300 to 6700 Å. The Balmer (up to H$\gamma$) and the strong helium lines have P Cygni-type profiles indicating a strong stellar wind. The H$\alpha$ emission, combined with spectral type B, usually classifies a star as Be, but for a Be star the luminosity class should be in the range V-III (cf. Underhill & Doazan 1982). Hammerschlag et al. (1979) and Parkes et al. (1980) classified the optical spectrum of Wray 977, using line strength ratios of Si IV/He I, Mg II/He I and O II/He I. The great strength of the O II lines, particularly O II 4349 Å, is indicative of a high luminosity. Therefore, they proposed that the spectral type of Wray 977 is B2 Ia.

**Table 1.** Log of observations of Wray 977 obtained with the ESO 3.6m telscope and CASPEC; the orbital phase is calculated from $\phi_0$ = JD2443906.56 mod(41.508) given in Sato et al. 1986

| JD-2440000 (Feb 1984) | $\Delta\lambda$ (Å) | $t_{exp}$ (sec.) | Orbital phase |
|---|---|---|---|
| 5743.842 | 4300-5100 | 3000 | 0.263 |
| 5744.850 | 5500-6000 | 600 | 0.288 |
| 5746.878 | 6500-6700 | 900 | 0.336 |

In the atlas of optical spectra of galactic B supergiants (Lennon et al. 1992), only one of the early B supergiants shows emission in the Balmer series: HD190603 with spectral type B1.5 Ia+. The '+'-classification refers to a class of "super-supergiants" (Keenan 1971), initially consisting of the four reddest of the brightest stars in the Large Magellanic Cloud (Feast & Thackeray 1956). Van Gen-deren (1979) named these stars hypergiants. A well-known early-B hypergiant is $\zeta^1$ Sco (HD 152236), one of the most-luminous stars in our galaxy. It is located in the neighbourhood of the open cluster NGC 6231 in the association I Scorpii, which has a distance modulus $(m - M)_0 = 11.5$ (Stothers 1972). The spectral type of $\zeta^1$ Sco is B1.5 Ia+, $T_{eff} = 19,000$ K (Appenzeller & Wolf 1979), and the absolute magnitude $M_V = -8.74$ (Humphreys 1978). The ultraviolet P Cygni profiles can be found in Wolf & Appenzeller (1979) and are comprised in the new IUE atlas of Snow et al. (1994).

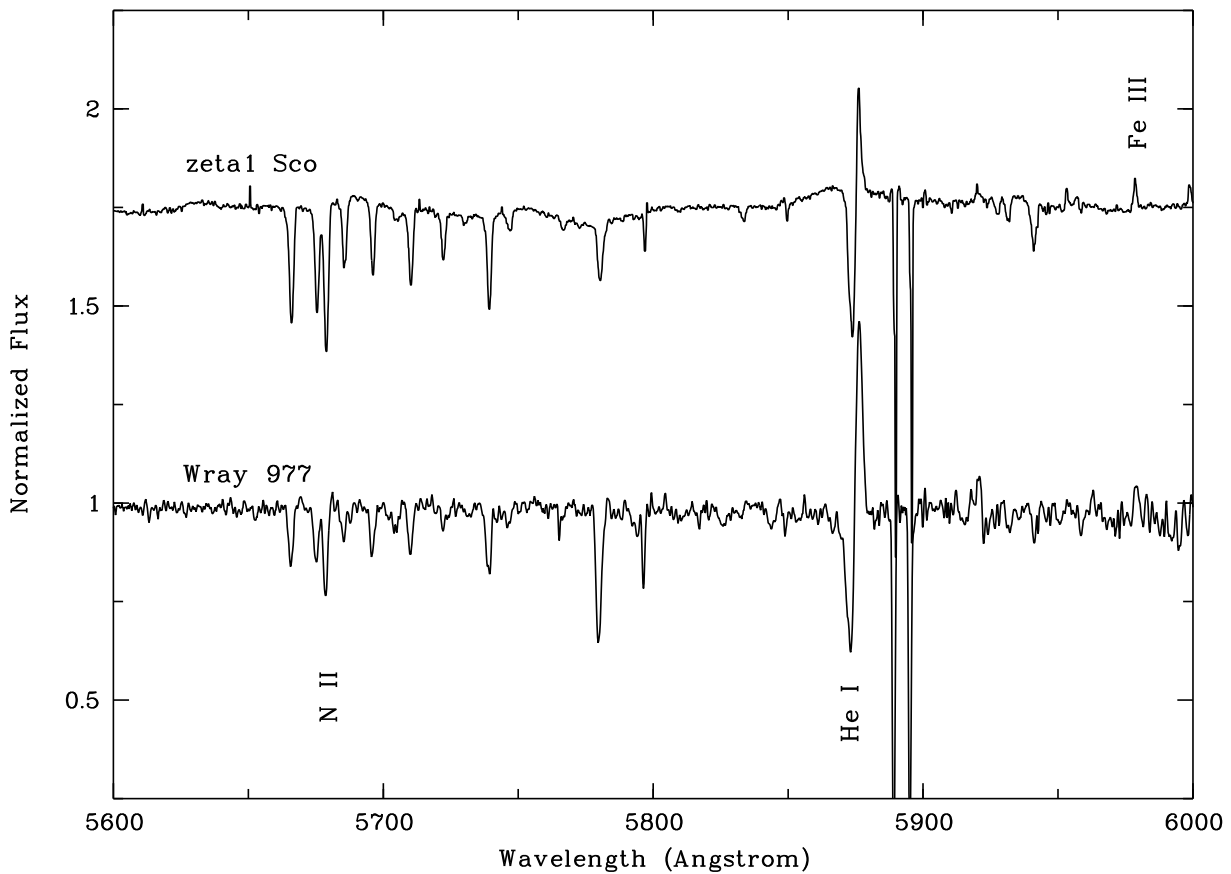

**Fig. 2.** The optical spectrum of $\zeta^1$ Sco and Wray 977 in the wavelength region from 5600 to 6000 Å. Note the weaker N II lines in the spectrum of Wray 977 and the stronger emission in the He I profile at 5876 Å. Although a fluorescence line of Fe III is present in the spectrum of $\zeta^1$ Sco, the bad S/N prohibits a firm detection of this line in this part of the spectrum of Wray 977

If we compare the optical spectrum of Wray 977 with that of $\zeta^1$ Sco (Figs. 1 and 2), we note that the two spectra are very similar. To enable a more quantitative comparison, we have measured the equivalent widths (EWs) of the most important spectral lines in the wavelength range 4300-6700 Å (Table 2). The uncertainty in EW (due to e.g. the normalization of the spectra) is estimated to be 10% or less. A careful inspection of the spectra shows that the O II lines in the spectrum of Wray 977 are stronger and the N II lines weaker than their counterparts in the spectrum of $\zeta^1$ Sco, which is confirmed by the measured EWs listed in Table 2. Using the atlases of OB-star spectra by Lennon et al. (1992) and Walborn & Fitzpatrick (1990), we conclude that Wray 977 has a slightly earlier spectral type than $\zeta^1$ Sco: B1 in stead of B1.5. For this spectral type the relative strength of the He I 4387 Å line compared to the Si III lines at $\lambda\lambda 4552$-68-75 provides a good indication of the luminosity class. The fact that the He I line is *weaker* than the Si III lines clearly demonstrates that the luminosity class is Ia+ (see Walborn & Fitzpatrick 1990). Indeed, the spectrum of HD 169454 (B1Ia+) resembles the spectrum of Wray 977 even more (Wolf, private communication).

Wolf & Stahl (1985) found that fluorescence lines of Fe III are sensitive to luminosity in early-B supergiants. An example of an Fe III emission line at 5979 Å (and at 5999 Å at the border of the plot) in the spectrum of $\zeta^1$ Sco is shown in Fig. 2, in agreement with the hypergiant nature of this star. Due to the bad signal-to-noise in this presence of this line, although an emission feature might be identified at 5979 Å.

**Table 2.** Line equivalent widths for $\zeta^1$ Sco and Wray 977. The uncertainty in EW is estimated to be 10% or less. Some unidentified lines may correspond to interstellar features

| Line | $\lambda$ (Å) | EW (Å) ($\zeta^1$ Sco) | EW (Å) (Wray 977) | Remarks |
|---|---|---|---|---|
| O II (2) | 4319 | 0.33 | 0.66 | blend |
| H$\gamma$ | 4340 | 0.99 | 0.50 | |
| O II (2) | 4346 | 0.27 | 0.34 | |
| O II (2) | 4349 | 0.58 | 0.76 | |
| O II (2) | 4367 | 0.28 | 0.23 | |
| He$^1$ I | 4387 | 0.54 | 0.63 | |
| O II (5) | 4417 | 0.70 | 0.91 | blend |
| He$^1$ I | 4437 | 0.09 | 0.07 | |
| unident | 4446: | 0.22 | | |
| He$^3$ I | 4471 | 0.81 | 0.47 | |
| Mg II (4) | 4481 | 0.33 | 0.29 | |
| unident | 4502: | | 0.33 | interst.? |
| Si III (2) | 4553 | 0.69 | 0.84 | |
| Si III (2) | 4568 | 0.49 | 0.53 | |
| Si III (2) | 4575 | 0.37 | 0.41 | |
| O II (15) | 4591 | 0.28 | 0.28 | |
| O II (15) | 4596 | 0.20 | 0.31 | |
| N II (5) | 4601 | 0.36 | 0.24 | |
| N II (5) | 4607 | 0.39 | 0.28 | |
| N II (5) | 4614 | 0.26 | 0.21 | |
| N II (5) | 4621 | 0.30 | 0.21 | |
| N II (5) | 4631 | 0.54 | 0.42 | |
| O II (1) | 4642 | 0.85 | 0.92 | blend |
| O II (1) | 4649 | 0.85 | 1.23 | blend |
| O II (1) | 4662 | 0.25 | 0.40 | |
| unident | 4668: | | 0.27 | |
| O II (1) | 4676 | 0.37 | 0.90 | blend |
| unident | 4704: | 0.13 | 0.33 | blend |
| He$^3$ I | 4713 | 0.30 | 0.46 | blend |
| unident | 4726: | | 0.25 | interst.? |
| Si III (9) | 4820 | 0.19 | 0.12 | |
| Si III (9) | 4829 | 0.16 | 0.18 | |
| H$\beta$ | 4861 | 0.26 | -1.13 | |
| He$^1$ I | 4921 | 0.77 | 0.60 | |
| N II (24) | 4994 | 0.09 | 0.11 | |
| N II (19) | 5001 | 0.48 | 0.32 | blend |
| He$^1$ I | 5015 | 0.36 | -0.17 | blend |
| He$^1$ I | 5047 | 0.75 | 0.61 | blend |
| N II (3) | 5667 | 0.63 | 0.33 | |
| N II (3) | 5680 | 1.32 | 0.76 | blend |
| N II (3) | 5686 | 0.22 | 0.27 | blend |
| Al III (2) | 5696 | 0.35 | 0.32 | blend |
| N II (3) | 5711 | 0.39 | 0.32 | |
| Al III (2) | 5723 | 0.24 | 0.25 | |
| Si III (4) | 5740 | 0.52 | 0.49 | |
| Fe I ? | 5781 | 0.47 | 0.89 | blend |
| He$^3$ I | 5876 | 0.63 | 0.13 | |
| Na I | 5890 | 0.78 | 1.18 | interst. |
| Na I | 5896 | 0.73 | 1.13 | interst. |
| H$\alpha$ | 6563 | -5.16 | -7.10 | |
| He$^1$ I | 6678 | 1.08 | 0.60 | |

Finally, the P Cygni shape of the H Balmer and the strong He I lines is consistent with the extreme luminosity of Wray 977. Rosendhal (1973) showed that the net strength of H$\alpha$ is correlated with luminosity for early-type supergiants. Wolff & Tully (1984) derived a similar relation between H$\alpha$ strength and $L$, but then for late-B supergiants. The H$\alpha$ equivalent width ($-7.1$ and $-5.2$ for Wray 977 and $\zeta^1$ Sco, respectively) suggests that the luminosity of Wray 977 is even higher than that of $\zeta^1$ Sco. Regarding the small number of early-type hypergiants and their uncertain absolute magnitudes, we adopt for Wray 977 the same luminosity as $\zeta^1$ Sco, although this might be a lower limit.

From the above, we conclude that the spectral type of Wray 977 is B1 Ia+ and that its absolute magnitude is $M_V = -8.7$. Then, using BC = $-1.87$ (Schmidt-Kaler 1982) and E(B$-$V) = 1.9 (Van Dishoeck & Black 1989), we find that the luminosity of Wray 977 is $L = 1.3 \times 10^6 L_\odot$ and its distance $d = 5.3$ kpc, which is a three-fold increase over the current value. Consequently, the X-ray luminosity of GX301-2 is an order of magnitude higher, i.e. in the range between $6.6 \times 10^{35}$ and $7.3 \times 10^{37}$ erg s$^{-1}$ (Van Paradijs 1991).

## 3. Wray 977 (GX301-2) revisited

The new classification of Wray 977 makes it the first hypergiant for which a lower limit on the mass is known. We will argue below that the lower mass limit has to be increased. In the previous section we introduced new values for the luminosity of and the distance to Wray 977. Combined with the adopted effective temperature of 20,800 K corresponding to spectral type B1 (Schmidt-Kaler 1982), we derive for the radius $R = 87 R_\odot$ (previously 43$R_\odot$, Parkes et al. 1980). This has important consequences for the binary system. The projected semi-major axis of the X-ray pulsar follows from pulse-timing analysis: $a_X \sin i = 159 \pm 1.5 \, R_\odot$ (Sato et al. 1986). The derived mass function of the system is:

$$f(m) = \frac{(M_{HG} \sin i)^3}{(M_X + M_{HG})^2} = 31.9 \pm 0.8 M_\odot \quad (1)$$

with $i$ the system's inclination, and M$_{HG}$ and M$_X$ the masses of the hypergiant and the X-ray pulsar, respectively. If we define the mass ratio $q = M_X/M_{HG}$, we can write Eq. 1 as:

$$M_{HG} = \frac{31.9(q+1)^2}{(\sin i)^3} \quad (2)$$

The radius of the hypergiant and the absence of X-ray eclipses set an upper limit to the system's inclination. For a radius of 43$R_\odot$, eccentricity $e = 0.47$, and periastron we can confirm. With $R = 87 R_\odot$, however, we derive $i \leq 64°$ (for $0.02 \leq q \leq 0.06$). From Eq. 2 it then follows that the lower limit for the mass of Wray 977 is $M_{HG} \geq 48$ $M_\odot$. According to evolutionary calculations including mass loss (Schaller et al. 1992) and given the effective temperature and luminosity of Wray 977, its "theoretical" mass is 52 $M_\odot$ ($M_{ZAMS} = 65$ $M_\odot$) if it is a post main-sequence star, or 33 $M_\odot$ ($M_{ZAMS} = 54$ $M_\odot$) if it is a post red supergiant. The new lower limit on the mass is consistent with Wray 977 being a post main-sequence star with a mass close to 50 $M_\odot$.

*3.1. Stellar-wind accretion onto GX301-2*

All B1 hypergiants are known to show a progression in radial velocity in the Balmer series with a range of about 200 km s$^{-1}$ between H$\alpha$ and members of the Balmer series having an upper quantum number $n \geq 9$ (Sterken & Wolf 1978). This Balmer progression is also observed for Wray 977 (Parkes et al. 1980), and even the strong He I lines indicate acceleration of the atmosphere. The terminal velocity of the wind is normally deduced from UV resonance lines, but due to the large interstellar extinction no UV spectra of Wray 977 are available. On the other hand, UV spectra of $\zeta^1$ Sco give $v_\infty = 390$ km s$^{-1}$ (Prinja et al. 1990) and similar values are obtained for other B1 hypergiants.

The mass-loss rate $\dot{M}$ of $\zeta^1$ Sco has been determined from radio observations to be $10^{-5}$ $M_\odot$ yr$^{-1}$ (Bieging et al. 1989). The relation between mass loss and $T_{eff}$ and $L$ of Lamers and Leitherer (1993), based on a sample of O stars, predicts $\dot{M} = 1.7 \times 10^{-5}$ $M_\odot$ yr$^{-1}$, given the $T_{eff}$ and $L$ of Wray 977. In Fig. 3 we show the H$\alpha$ profile of Wray 977 and three theoretical profiles resulting from the SEIBALMER program of Lamers (1994). Using departure coefficients that result from the radiative transfer code of De Koter (1993) for a B1 supergiant, the H$\alpha$ profile is computed for two different mass-loss rates: $10^{-5}$ (triangles) and $5 \times 10^{-6}$ $M_\odot$ yr$^{-1}$ (diamonds). The squares give the results for LTE populations and $\dot{M} = 10^{-5}$ $M_\odot$ yr$^{-1}$. For all models $v_\infty$ was set to 400 km s$^{-1}$ and the $\beta$-parameter of the velocity law is equal to one. Although none of the models can fit the blue-shifted P Cygni absorption, we estimate that the mass-loss rate derived from the H$\alpha$ profile is between $5 \times 10^{-6}$ and $10^{-5}$ $M_\odot$ yr$^{-1}$.

The X-ray luminosity produced by wind accretion can be estimated using the formalism of Bondi & Hoyle (1944). The accretion radius is given by $r_{acc} = 2GM_X v_{rel}^{-2}$ and the resulting X-ray luminosity is

$$L_X = \dot{M} \frac{\pi r_{acc}^2}{4\pi r^2} \frac{GM_X}{R_X}. \qquad (3)$$

The large eccentricity of the orbit ($e = 0.47$) implies that the separation between the binary components and the orbital velocity of the neutron star strongly depend on radius in periastron is of the same order of magnitude as in apastron (3.7 and 6.2 $R_\odot$, respectively). This is caused by the fact that the relative velocity (i.e. the vector sum of orbital and wind velocity) does not change much as a function of orbital phase. The orbital velocity is at its maximum and the wind velocity at its minimum in periastron, and vice versa in apastron. Using the above derived values for the system parameters, we find that the X-ray luminosity is $4.1 \times 10^{37}$ and $1.0 \times 10^{37}$ erg s$^{-1}$ in periastron and apastron, respectively. This is compatible with the observed "average" X-ray luminosity of $\sim 10^{37}$ erg s$^{-1}$ (see section 2), but does not explain by itself the large variations in X-ray flux (e.g. Haberl 1991). We note that periastron passage is very close to the star, of the order of 10 $R_\odot$ above the stellar surface.

## 4. Discussion

There is no consensus about the evolutionary status of early B hypergiants; it is suggested that these very luminous stars are the progenitors of Luminous Blue Variables (LBVs). In fact, Wray 977 has a spectral type and luminosity similar to that of P Cygni, but the mass-loss features in the spectra of e.g. P Cygni and S Dor are more extreme (cf. Stahl & Wolf 1982). LBVs are found close to the Humpreys-Davidson limit in the HRD and are characterized by phases of extreme mass loss where $\dot{M}$ can reach values of $10^{-4} - 10^{-3}$ $M_\odot$ yr$^{-1}$ (Lamers 1989). Lamers & Fitzpatrick (1988) proposed that the LBV phase is related to the fact that these massive stars reach their Eddington limit while they evolve towards the red. As a consequence, they lose their outer stellar layers. After the LBV phase, the blue supergiant will become a Wolf-Rayet star. As the name suggests, LBVs exhibit dramatic variability in their optical light; Wray 977 shows variations of $\sim 0.1$ mag in V (e.g. Pakull 1982) like other early-type hypergiants.

The occurrence of an LBV phase in massive binaries might explain the existence of some double neutron star systems (like PSR 2303+46) that cannot have been formed by common-envelope evolution (Van den Heuvel et al. 1994). In this scenario, the energy required to expel the envelope of the supergiant (i.e. the progenitor of the second neutron star) is derived from the orbital energy of the neutron star companion. If one applies the formalism of CE-evolution to the system PSR 2303+46, the orbital period of 12.33d implies that the progenitor system had an orbital period $\geq 10.8$yr, which excludes the occurrence of a spiral-in phase. If a star spontaneously sheds its envelope without the need of a companion spiralling in, this problem can be overcome. It appears that, if the progenitor of the PSR 2303+46 system was a HMXB with a supergiant that undergoes an LBV phase ($M_{ZAMS} \gtrsim 40 - 45$ $M_\odot$) with a high mass-loss rate, a consistent picture for its origin can be found.

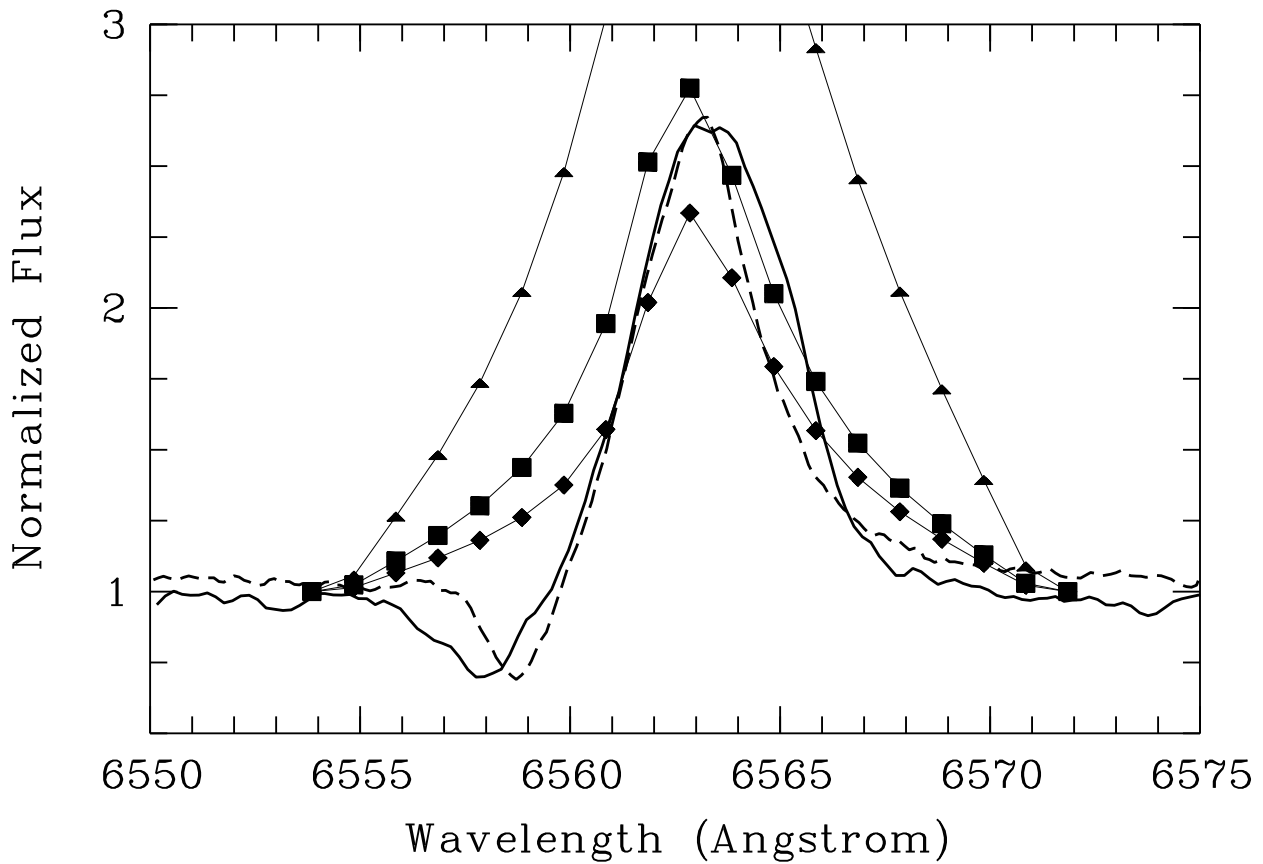

**Fig. 3.** The H$\alpha$ line of Wray 977 (thick line) and $\zeta^1$ Sco (dashed line) and three model calculations from the SEIBALMER-method (Lamers 1994). The triangles and diamonds, respectively, correspond to a model with a mass-loss rate of $10^{-5}$ and $5 \times 10^{-6}$ M$_\odot$ yr$^{-1}$, $v_\infty = 400$ km s$^{-1}$, and velocity law parameter $\beta = 1$; The squares represent a model with $\dot{M} = 10^{-5}$ M$_\odot$ yr$^{-1}$, $v_\infty = 400$ km s$^{-1}$, $\beta = 1$, but LTE populations

The new lower limit on the mass of Wray 977 has consequenses for the observed lower mass limit for blackhole formation in a binary as derived by Van den Heuvel & Habets (1984), based on the fact that GX301-2 is a neutron star. We propose that this limit is shifted to $M_{ZAMS} \gtrsim 50$ M$_\odot$. Since it is not clear if all stars with an initial mass exceeding 50 M$_\odot$ will indeed form a black hole (see e.g. Maeder 1992), the newly determined lower mass limit might have important consequences for the number of black holes formed in binaries.

*Acknowledgements.* We thank Dr. O. Stahl for providing us with an optical spectrum of $\zeta^1$ Sco and Dr. A. van Genderen for his information about hypergiants. Dr. D. Lennon's help in classifying the optical spectrum of Wray 977 is highly appreciated. Dr. T. Bedding and Dr. G. Thimm are acknowledged for their helpful suggestions. We are grateful to the referee Dr. Wolf to show the similarity between the spectra of HD169454 and Wray 977. ER is a postdoctoral researcher of the Belgian National Fund for Scientific Research.